\documentclass[cameraready]{Interspeech}

\title{Beyond Short Segments : Expanding Speaker Embeddings\\with Vector Archives}

\author[orcid={0009-0000-3470-5381}, equalcontribution]{Hyunku}{Kang}
\author[orcid={0009-0003-2010-2004}, equalcontribution]{Minkyu}{Cho}
\author[orcid={0000-0003-4085-2470}, correspondingauthor]{Chanwoo}{Kim}

\address{Korea University, Republic of Korea}

\email{\{kahk000,minkyucho,chanwcom\}@korea.ac.kr}


\usepackage{comment}

\usepackage{amsmath,graphicx,hyperref}
\usepackage{booktabs}
\usepackage{array}
\usepackage{graphicx}
\usepackage{float}
\usepackage{algorithm}
\usepackage{algpseudocode}
\usepackage{subcaption}

\begin{document}

\maketitle
\footnotetext{The source code is publicly available at: \url{https://github.com/slp-lab-research/vam_ecapa}}

\begin{abstract}
The performance of state-of-the-art speaker verification (SV) systems severely degrades on short utterances due to insufficient speaker-specific information. To address this critical challenge, we propose the Vector Archive Mapping ECAPA (VAM-ECAPA), a novel system designed to enhance feature extraction from short-duration speech. The core of our system is the Transformer-based Vector Archive Mapping with Statistical Pooling (TVAMSP) module, which enriches information-scarce features by mapping them against a learnable Vector Archive of canonical speaker traits. By integrating the TVAMSP module into a strong WavLM+ECAPA-TDNN baseline, our system learns to map sparse features from short segments into robust, discriminative speaker representations. Experiments on the VoxCeleb1 benchmark show that our proposed VAM-ECAPA achieves a highly competitive EER of 8.334\% on 1-second test segments, a 54.8\% relative error reduction compared to a conventionally-trained baseline.
\end{abstract}

\section{Introduction}

Speaker verification (SV) has seen significant progress, with deep neural network based systems achieving highly competitive performance on standard benchmarks \cite{intro1, new3,add1}. A key point of this advancement is the adoption of self-supervised learning (SSL), where large-scale pre-trained models like WavLM \cite{wavlm} learn robust and generalizable speech representations from vast amounts of unlabeled data \cite{wavlm, new6, new7, eunseo}. These SSL-based backbones, when fine-tuned for SV, have become a foundational approach for building highly competitive systems \cite{superb, new8}.

Despite these advances, a critical challenge persists: the performance of even the most powerful SV systems degrades dramatically when verifying speakers from very short utterances, typically those under three seconds \cite{intro4}. This performance drop occurs because short segments often lack the rich coarticulatory cues and prosodic contours that unfold over longer durations, which are essential for extracting a stable and discriminative speaker embedding \cite{intro5, new4}.  This limitation directly hinders real-world deployment. In voice-activated devices, a user command typically lasts under two seconds. In phone-based authentication systems, the available speech is often one to three seconds. In these settings, a system that requires longer input is simply not usable. The gap between laboratory performance on long utterances and real deployment conditions
on short ones represents one of the most pressing open problems in speaker verification.

To bridge this performance gap, we argue that simply finetuning a large SSL model is not enough for these challenging conditions \cite{intro6, new9}. Prior work has addressed this problem by aggregating multiple short segments or applying meta-learning strategies \cite{intro7,intro9}. However, these methods either require more than one utterance at inference time or do not directly enrich frame-level features. We take a different approach by enriching each frame through a learnable set of canonical speaker traits \cite{new12}.

This set, which we call the Vector Archive, is trained end-to-end to represent the feature space of longer utterances. By mapping short-utterance features against this archive, we introduce Vector Archive Mapping ECAPA (VAM-ECAPA), an end-to-end system that recovers missing speaker information without extra input at inference time.

\section{Related Work}
Short utterance speaker verification has been studied from several angles \cite{new1,new2}. One line of work focuses on aggregating multiple short segments at inference time to build a richer representation. While effective, these methods require more than one utterance, which limits their use in real-time scenarios. Another line of work applies meta-learning to train models that generalize better across utterance lengths \cite{intro7,new11}. These approaches improve robustness but do not directly address the sparse information in individual frames. Data augmentation strategies such as duration-based sampling have also been explored to expose models to short segments during training, yet they do not enrich the features themselves \cite{met1}. SSL backbones have raised the performance ceiling for speaker verification. Wav2Vec 2.0, HuBERT, and WavLM have each been applied to SV with strong results, but their representations still degrade on short inputs because the underlying frame-level features remain sparse. 

All of these directions share a common limitation: they improve robustness to short utterances without directly enriching frame-level features at extraction time. This gap is what our work targets. Our approach is also related to memory-augmented neural networks, where external or learnable memory representations complement input features \cite{final1,final2}. Similar memory mechanisms have also been explored in speaker verification and diarization \cite{final3,final4}. Unlike prior memory-based approaches, VAM uses a compact set of learnable vector archives specifically designed to compensate for information scarcity in short-duration speaker verification.

\section{Proposed Methods}
\label{sec:pagestyle}

Our proposed model, VAM-ECAPA, addresses information scarcity in short utterances through a three-stage pipeline, as illustrated in Fig. 1(a). First, a pre-trained WavLM backbone serves as a universal feature extractor. Second, our proposed TVAMSP module enhances the features to be rich in speaker-specific information. Finally, a standard ECAPA-TDNN model \cite{ecapa} acts as a backend encoder.


\begin{figure}[!h]
    \centering
    \begin{subfigure}[b]{0.48\columnwidth}
        \centering
        \includegraphics[width=\textwidth]{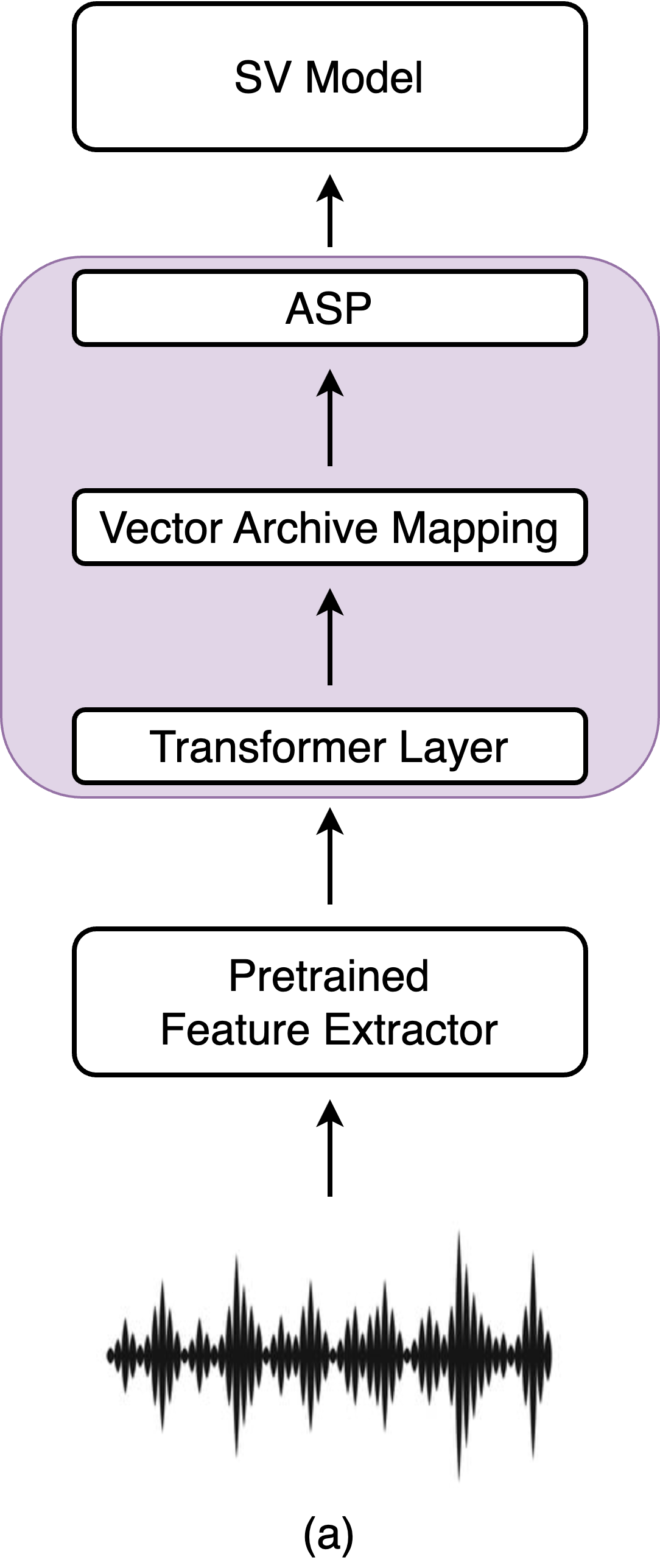}
        \caption{Overall VAM-ECAPA}
        \label{fig:pipeline}
    \end{subfigure}
    \hfill 
    \begin{subfigure}[b]{0.48\columnwidth}
        \centering
        \includegraphics[width=\textwidth]{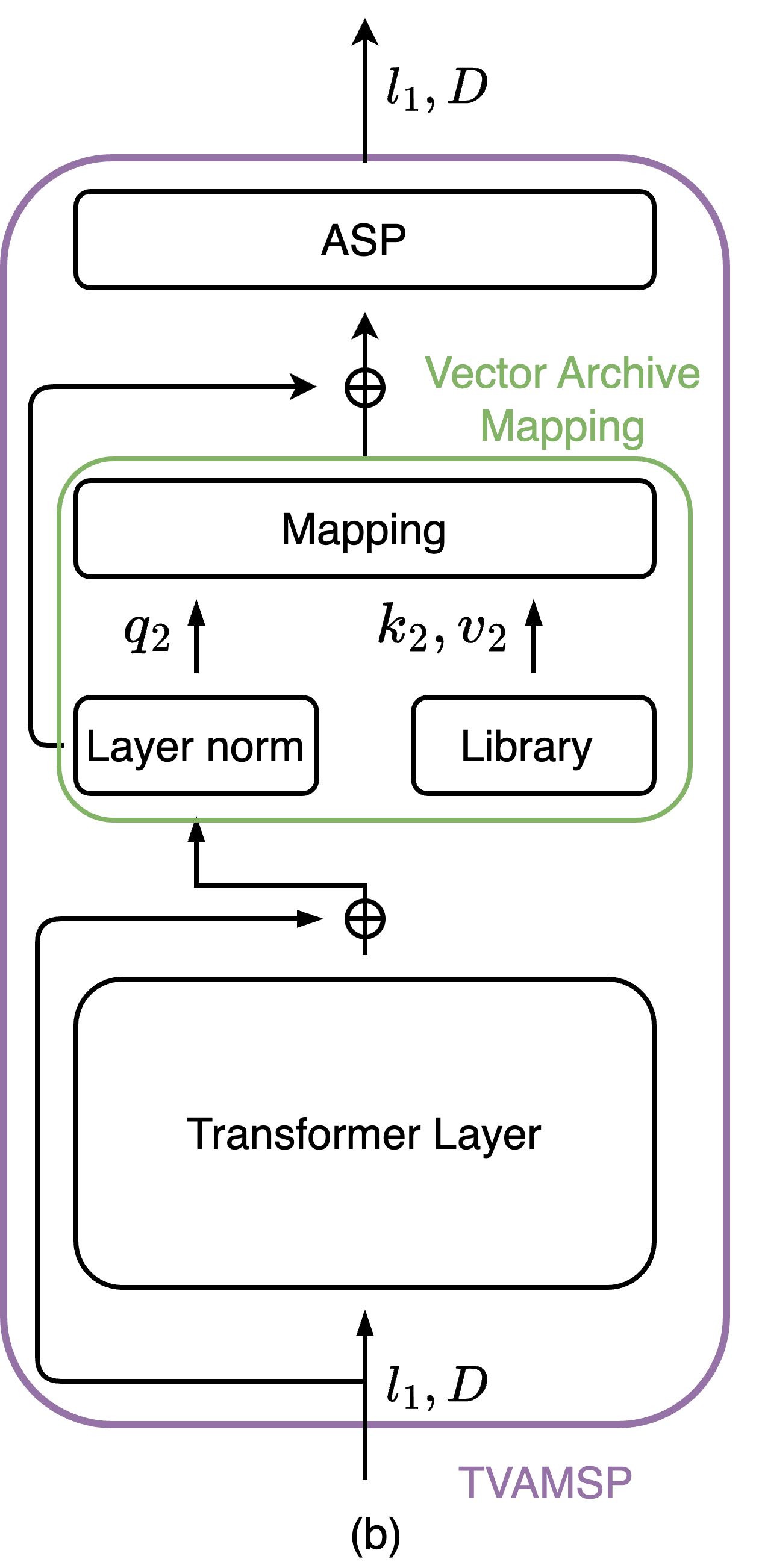}
        \caption{TVAMSP Module}
        \label{fig:module}
    \end{subfigure}
    \caption{Proposed Vector Archive Mapping ECAPA-TDNN (VAM-ECAPA) architecture for short-utterance speaker verification. 
    \textbf{(a)} Overall pipeline, where the Transformer-based Vector Archive Mapping with Statistical Pooling (TVAMSP) module refines WavLM features before the ECAPA-TDNN speaker encoder. 
    \textbf{(b)} TVAMSP module, consisting of a Transformer layer, Vector Archive Mapping, and Attentive Statistics Pooling (ASP).}
    \label{fig:overall_architecture}
\end{figure}

\subsection{Feature Extraction with WavLM}
The first stage of our pipeline is to extract a strong, general-purpose speech representation. We employ the pre-trained WavLM as our backbone, as it has demonstrated exceptional robustness across diverse domains, providing a strong foundation for feature extraction even in challenging scenarios \cite{wavlm}. Following the SUPERB \cite{superb} methodology, we compute a weighted sum of all layer representations to generate a single feature sequence $f \in \mathbb{R}^{l_1 \times D}$.

\subsection{Transformer-based Vector Archive Mapping with Statistics Pooling (TVAMSP)}
The TVAMSP module is the core of our system, designed to transform the general features $f$ into representations tailored for short utterance SV. As shown in Fig.~\ref{fig:module}, it consists of three sequential sub-components.
\begin{figure}[!h]
  \centering
  \includegraphics[width=0.9\linewidth]{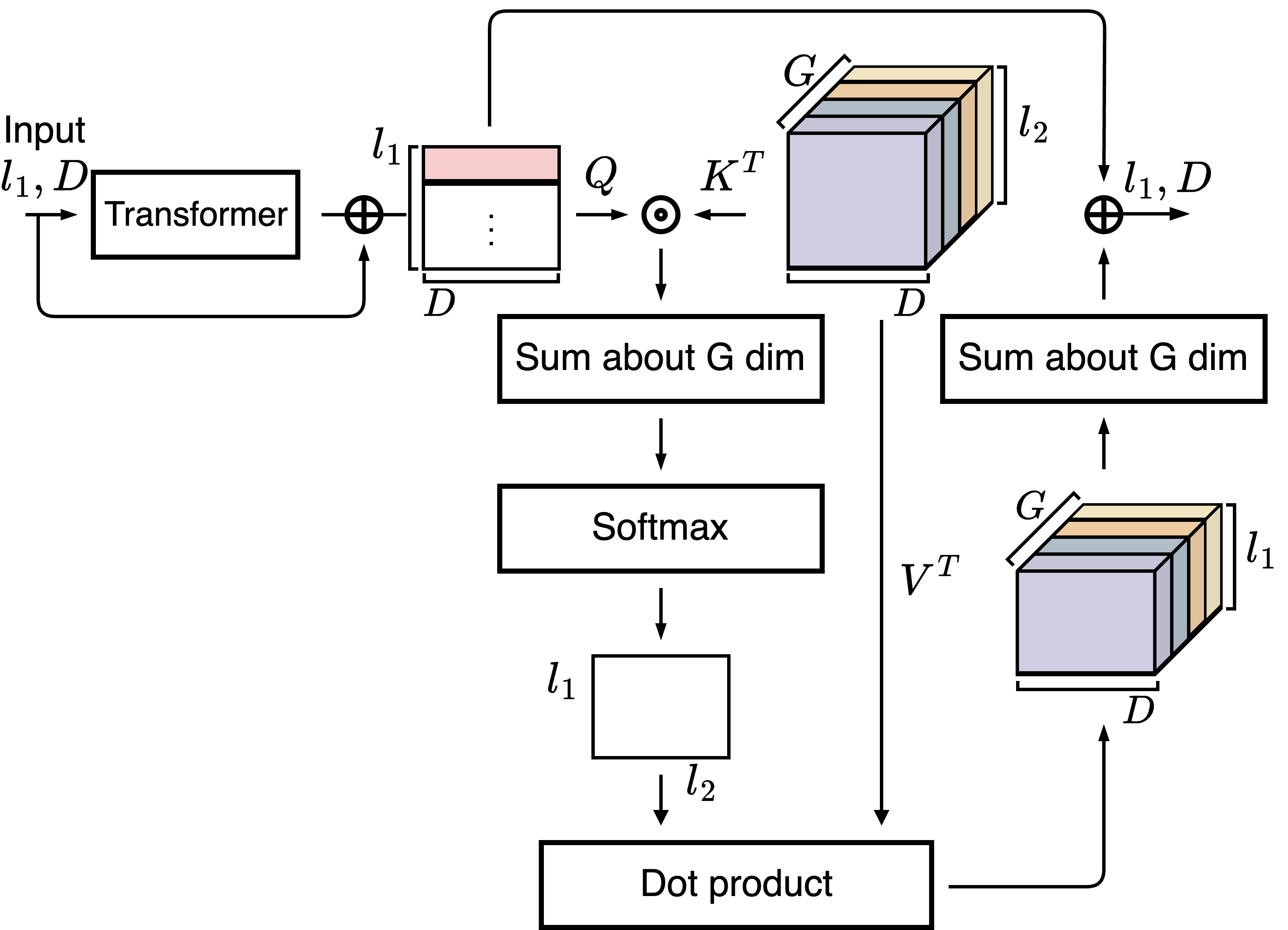}
  \caption{Detailed mechanism of Vector Archive Mapping. The input sequence $O_1$ is transformed by referencing a learnable Library with $G$ Archives, each containing $l_2$ vectors.}
  \label{fig:vam_mechanism}
\end{figure}

\subsubsection{Feature Enhancement with Transformer Layer}
While the WavLM features are powerful, they represent each frame independently \cite{met2}. To explicitly model the temporal dependencies within an utterance, we first process the features $f \in \mathbb{R}^{l_1 \times D}$ with a standard Transformer layer \cite{met6}. This self-attention mechanism captures relationships across frames and produces a context-aware representation $O_1$ for the subsequent mapping stage.

\begin{table*}[t]
\centering
\caption{Short-utterance speaker verification results on VoxCeleb1. Vox1-O, Vox1-E, and Vox1-H denote the original, extended, and hard trial lists. ASV, EER, and MinDCF denote automatic speaker verification, Equal Error Rate, and Minimum Detection Cost Function, respectively.}
\renewcommand{\arraystretch}{1.5}
\setlength{\tabcolsep}{2.0pt}
\resizebox{\textwidth}{!}{
\begin{tabular}{ll||cc|cc|cc||cc|cc|cc||cc|cc|cc|}
\hline
\textbf{Encoder} & \textbf{ASV} & \multicolumn{6}{c||}{\textbf{Vox1-O}} & \multicolumn{6}{c||}{\textbf{Vox1-E}} & \multicolumn{6}{c|}{\textbf{Vox1-H}} \\ \cline{3-20}
                 &              & \multicolumn{2}{c|}{\textbf{3s}} & \multicolumn{2}{c|}{\textbf{2s}} & \multicolumn{2}{c||}{\textbf{1s}} & \multicolumn{2}{c|}{\textbf{3s}} & \multicolumn{2}{c|}{\textbf{2s}} & \multicolumn{2}{c||}{\textbf{1s}} & \multicolumn{2}{c|}{\textbf{3s}} & \multicolumn{2}{c|}{\textbf{2s}} & \multicolumn{2}{c|}{\textbf{1s}} \\ \cline{3-20}
                 &              & \textbf{EER} & \textbf{MinDCF} & \textbf{EER} & \textbf{MinDCF} & \textbf{EER} & \textbf{MinDCF} & \textbf{EER} & \textbf{MinDCF} & \textbf{EER} & \textbf{MinDCF} & \textbf{EER} & \textbf{MinDCF} & \textbf{EER} & \textbf{MinDCF} & \textbf{EER} & \textbf{MinDCF} & \textbf{EER} & \textbf{MinDCF} \\ \hline
Wav2vec 2.0      & ECAPA-TDNN   & 2.968 & 0.212 & 6.202 & 0.373 & 19.009 & 0.791 & 2.883 & 0.177 & 5.934 & 0.331 & 18.214 & 0.770 & 4.960 & 0.304 & 8.535 & 0.484 & 20.927 & 0.862 \\
HuBERT           & ECAPA-TDNN   & 2.760 & 0.186 & 5.959 & 0.346 & 19.326 & 0.797 & 2.734 & 0.166 & 5.773 & 0.314 & 18.569 & 0.768 & 4.666 & 0.296 & 8.334 & 0.470 & 21.045 & 0.860 \\
WavLM            & ECAPA-TDNN   & \textbf{2.393} & \textbf{0.169} & 5.242 & 0.366 & 18.437 & 0.780 & \textbf{2.462} & \textbf{0.150} & 5.295 & \textbf{0.295} & 18.059 & 0.746 & \textbf{4.395} & \textbf{0.283} & \textbf{7.877} & \textbf{0.460} & 20.449 & 0.842 \\
WavLM            & VAM-ECAPA (Ours) & 3.185  & 0.254     & \textbf{4.175}     & \textbf{0.335}     & \textbf{8.334} & \textbf{0.536}  & 3.541   & 0.265  & \textbf{4.512} & 0.332 & \textbf{8.511} & \textbf{0.551}  & 7.178  & 0.436 & 8.941 & 0.525 & \textbf{14.571}        & \textbf{0.746}           \\ \hline
\end{tabular}}
\label{tab:short-sequence}
\end{table*}

\subsubsection{Vector Archive Mapping}
This sub-component is our key contribution for mitigating information scarcity. The mechanism is related to cross-attention, but differs in a fundamental way. In standard cross-attention, the Keys and Values come from another input sequence, so the quality of the reference depends entirely on what is available at that moment. In our design, the Keys and Values are derived from a learnable Library trained end-to-end as a fixed model parameter. It encodes stable, canonical speaker traits from the training data, so even when an input utterance is very short, the model always has access to a rich and consistent reference.

This archive is implemented as a \textbf{Library} $L \in \mathbb{R}^{G \times l_2 \times D}$, composed of $G$ distinct \textbf{Archives}. The input features $O_1$ are projected to Queries ($Q \in \mathbb{R}^{l_1 \times D}$), while the Library $L$ is projected to Keys ($K \in \mathbb{R}^{G \times l_2 \times D}$) and Values ($V \in \mathbb{R}^{G \times l_2 \times D}$). First, we perform archive aggregation by summing the Keys and Values across all $G$ archives to create a single, unified reference set:
\begin{align}
    K_{\text{agg}} &= \sum_{g=1}^{G} K_{g,:,:} \; , \ V_{\text{agg}} = \sum_{g=1}^{G} V_{g,:,:}
\end{align}
The mapping scores are then computed via matrix multiplication and normalized, as shown below:

\begin{equation}
    A = Q K_{\text{agg}}^T \quad ; \quad \hat{A} = \operatorname{softmax}(A / \tau)
\end{equation}
These scores are then used to attend to the aggregated Values, producing the final enriched representation $O_2 = \hat{A} V_{\text{agg}} + O_1$. For clarity, the element-wise computation for the mapping score $A_{i,j}$ (the score for the $i$-th input frame against the $j$-th archive concept) is as follows:
\begin{equation}
    A_{i,j} = \sum_{g=1}^{G} \sum_{d=1}^{D} Q_{i,d} \cdot K_{g,j,d}
\end{equation}

Similarly, the element-wise computation for each component of the final output vector $O_{2_{i,d}}$ is as follows:

\begin{equation}
    O_{2_{i,d}} = \left( \sum_{j=1}^{l_2} \hat{A}_{i,j} \left( \sum_{g=1}^{G} V_{g,j,d} \right) \right) + O_{1_{i,d}}
\end{equation}

Through this mapping, each input frame is aligned to the speaker traits stored in the Library. This allows the model to recover speaker-discriminative information that is missing in short utterances.


\subsubsection{Feature Augmentation with Attentive Statistics}
The final step of the TVAMSP module is to infuse global, utterance-level context into every frame of the enhanced feature sequence $O_2 \in \mathbb{R}^{l_1 \times D}$. This is achieved via a feature augmentation mechanism that uses Attentive Statistics Pooling (ASP) \cite{met3} as its core component.

First, the ASP layer calculates a single utterance-level summary vector from the entire sequence $O_2$. It computes attention weights $\alpha$ over the temporal dimension, which are then used to derive the weighted mean $\mu$ and standard deviation $\sigma$:
\begin{align}
    \alpha &= \operatorname{softmax}(W_2 \operatorname{tanh}(W_1 O_2^T)) \\
    \mu &= \sum_{t=1}^{l_1} \alpha_t O_{2_t} \quad ; \quad \sigma = \sqrt{\sum_{t=1}^{l_1} \alpha_t O_{2_t}^2 - \mu^2}
\end{align}
These statistics are concatenated and projected to form the final summary vector, formulated as $s = W_p[\mu \, ; \, \sigma]$. Instead of passing this vector directly to the next stage, we broadcast it across the temporal dimension and add it back to the frame-level sequence. This process enriches each individual frame with global information about the entire utterance, producing the final module output $O_3 = O_2 + s$. The resulting augmented feature sequence $O_3$ is then passed to the final speaker embedding encoder.

\subsection{Speaker Embedding Generation}
The utterance-level representation $O_3$ produced by the TVA\hspace{0pt}MSP module is then fed into an ECAPA-TDNN model, which serves as the final speaker embedding encoder. The powerful architecture of ECAPA-TDNN, with its SE-Res2Blocks \cite{new14,new15} and channel-dependent statistics, transforms $O_3$ into the final 192-dimensional speaker embedding $e$. This embedding is L2-normalized and scaled, making it suitable for robust similarity scoring using cosine distance in the speaker verification task.





\section{Experiments}


\subsection{Datasets and Metrics}
We follow the standard protocol for the VoxCeleb datasets \cite{exp1}, \cite{exp2} to ensure a fair and reproducible evaluation. Our models are trained exclusively on the large-scale development set of VoxCeleb2 to learn robust and generalizable speaker representations \cite{exp1}. Evaluation is performed on the official VoxCeleb1 test set to strictly assess the model's ability to generalize to unseen identities. We report performance on all three official trial lists (Vox1-O, E, and H) using Equal Error Rate (EER, $\%$) and Minimum Detection Cost Function (MinDCF) with $P_{\text{target}}=0.05$.

\subsection{Baseline and Training Details}
Our baseline system, pairing a WavLM backbone with an ECAPA-TDNN encoder, is trained using standard data augmentation \cite{exp4, exp5, lab1, lab2} and AAM Softmax loss \cite{add3, new13}. The training follows a three-stage recipe of initial head training, joint fine-tuning, and a final large-margin stage \cite{add2}. For our proposed VAM-ECAPA, the TVAMSP module's hyperparameters were motivated by the backbone's characteristics. Based on preliminary experiments showing the WavLM backbone yields the most stable features on segments of 3s or longer, we set the conceptual length of our Vector Archives to $l_2 = 149$, which corresponds to 3s of speech at the WavLM feature rate. This choice is intentional: the Archive is designed to represent the feature space that the backbone produces under stable conditions, so that short-utterance features can be mapped toward this more informative reference. The number of archives was empirically set to $G = 4$ for the best trade-off between performance and complexity.


\subsection{Baseline Performance Comparison}
To ensure our baseline is competitive, we first compare several state-of-the-art SSL backbones paired with an ECAPA-TDNN encoder on the full-length VoxCeleb1 test set. As shown in Table 2, the WavLM backbone consistently outperforms Wav2Vec 2.0 and HuBERT across all test conditions. This confirms its status as a top-tier feature extractor and justifies its selection as the foundation for our system. However, even this strongest backbone suffers severe degradation on short segments, motivating the need for the TVAMSP module evaluated next.


\begin{table}[h!]
\centering
\caption{Full-length VoxCeleb1 results using different self-supervised learning (SSL) backbones. Vox1-O, Vox1-E, and Vox1-H denote the original, extended, and hard trial lists. ASV, EER, and MinDCF denote automatic speaker verification, Equal Error Rate, and Minimum Detection Cost Function, respectively.}
\begingroup
\renewcommand{\arraystretch}{1.5}
\setlength{\tabcolsep}{2.0pt}
\resizebox{\columnwidth}{!}{%
\begin{tabular}{@{}llcccccc@{}}
\toprule
\textbf{Encoder} & \textbf{ASV} & \multicolumn{2}{c}{\textbf{Vox1-O}} & \multicolumn{2}{c}{\textbf{Vox1-E}} & \multicolumn{2}{c}{\textbf{Vox1-H}} \\
\cmidrule(r){3-4} \cmidrule(r){5-6} \cmidrule(r){7-8}
& & \textbf{EER(\%)} & \textbf{MinDCF} & \textbf{EER(\%)} & \textbf{MinDCF} & \textbf{EER(\%)} & \textbf{MinDCF} \\
\midrule
Wav2vec 2.0 & ECAPA-TDNN & 1.675 & 0.130 & 1.246 & 0.082 & 2.601 & 0.164 \\
HuBERT      & ECAPA-TDNN & 1.122 & \textbf{0.086} & 1.124 & 0.072 & 2.333 & 0.148 \\
WavLM       & ECAPA-TDNN & \textbf{0.973} & 0.087 & \textbf{1.032} & \textbf{0.069} & \textbf{2.243} & \textbf{0.145} \\
\bottomrule
\end{tabular}
}%
\endgroup
\label{tab:baseline_comparison}
\end{table}

\subsection{Short Utterance Verification Results}
Table 1 provides a comprehensive comparison across all three VoxCeleb1 trial lists at 3s, 2s, and 1s durations. Table 3 presents a more detailed breakdown on the Vox1-O test set, tracing the step-by-step contribution of each training recipe. As shown in Table 3, the standard baseline's performance degrades dramatically as duration decreases, with the EER rising from 2.393\% at 3s to 18.437\% at 1s. While adapting the ECAPA baseline to 1s training improves performance, reducing the EER from 18.437\% to 10.346\%, the comparison between the third and fourth rows of Table 3 isolates the effect of VAM under an identical training recipe. Under this controlled setting, adding VAM further reduces the EER from 10.346\% to 8.342\%, demonstrating that the gain is not solely due to short-segment retraining but also comes from the proposed archive-based feature compensation. 

In contrast, our proposed VAM-ECAPA is remarkably effective, achieving an EER of 8.334\% on the most challenging 1s segments, which represents a 54.8\% relative error reduction compared to the baseline. As shown in Table 1, this trend holds consistently across all three trial lists. On the harder Vox1-H condition, the baseline EER reaches 20.449\% at 1s, while our system reduces it to 14.571\%, a relative improvement of 28.7\%. The improvement is similarly consistent on Vox1-E, where the EER drops from 18.059\% to 8.511\% at 1s. These results confirm that the gains from our approach are not specific to one test condition but generalize across varying levels of trial difficulty. On 3s segments, VAM-ECAPA shows higher EER than the baseline across all trial lists. This outcome is expected and reflects a fundamental design principle of the TVAMSP module. The module is trained on 1s segments and learns to project sparse features toward a richer reference space. When the input is already 3s long, the WavLM backbone produces stable and information-rich features on its own. In this case, the mapping operation alters features that do not need correction, which introduces a small degree of distortion. This is not a flaw in the architecture but a direct consequence of optimizing for short-segment conditions. The system is designed for deployment scenarios where utterances are short, and it achieves its goal in exactly those conditions.

\begin{table}[h!]
\centering
\caption{Short-utterance results on Vox1-O, the original VoxCeleb1 trial list. EER and MinDCF denote Equal Error Rate and Minimum Detection Cost Function. pt, ft, lft, and w/o denote pre-training, fine-tuning, large-margin fine-tuning, and without, respectively.}
\renewcommand{\arraystretch}{1.5}
\setlength{\tabcolsep}{2.0pt}
\resizebox{\linewidth}{!}{
\begin{tabular}{@{}llcccccccc@{}}
\toprule
\textbf{SV System} & \textbf{Training Recipe} & \multicolumn{2}{c}{\textbf{3s}} & \multicolumn{2}{c}{\textbf{2s}} & \multicolumn{2}{c}{\textbf{1s}} \\ 
\cmidrule(lr){3-4} \cmidrule(lr){5-6} \cmidrule(lr){7-8}
                   &                         & \textbf{EER} & \textbf{MinDCF} & \textbf{EER} & \textbf{MinDCF} & \textbf{EER} & \textbf{MinDCF} \\ \midrule
\textbf{WavLM +}   & 3sec pt 0.2m +          &              &                 &              &                 &              &                 \\
\textbf{ECAPA-TDNN} & 3sec ft 0.2m +         & \textbf{2.393}        & \textbf{0.169}           & 5.242        & 0.366           & 18.437       & 0.780           \\
\textbf{(Baseline)} & 6sec  lft 0.4m           &              &                 &              &                 &              &                 \\ \midrule
\textbf{WavLM +}   & 1sec pt 0.2m +          &              &                 &              &                 &              &                 \\
\textbf{ECAPA-TDNN} & 1sec ft 0.2m +         & 4.094        & 0.307           & 5.606        & 0.408           & 11.931       & 0.667           \\
                   & w/o lft                 &              &                 &              &                 &              &                 \\ \midrule
\textbf{WavLM +} & 1sec pt 0.2m + &  &  &  &  &  &  \\
\textbf{ECAPA-TDNN} & w/o ft + & 3.169 & 0.243 & 4.648 & 0.347 & 10.346 & 0.606 \\
& w/o lft & &  &  &  &  &   \\ \midrule
\textbf{WavLM +} & 1sec pt 0.2m + &  &  &  &  &  &  \\
\textbf{VAM-ECAPA} & w/o ft + & 3.271 & 0.265 & 4.261 & 0.341 & 8.342 & 0.539 \\
\textbf{(Ours)} & w/o lft & &  &  &  &  &   \\ \midrule
\textbf{WavLM +}   & 1sec pt 0.1m +          &              &                 &              &                 &              &                 \\
\textbf{VAM-ECAPA} & w/o ft +                 &  3.185       & 0.254           &  \textbf{4.175}       & \textbf{0.335}                & \textbf{8.334}        & \textbf{0.536}           \\
\textbf{(Ours)}    & w/o lft                 &              &                 &              &                 &              &                 \\ 
\bottomrule
\end{tabular}
}
\label{tab:vox1o_performance}
\end{table}

\begin{table}[h!]
\centering
\caption{Ablation study of the Transformer-based Vector Archive Mapping with Statistical Pooling (TVAMSP) module on 1-second Vox1-O utterances. EER, VAM, ASP, and Res. denote Equal Error Rate, Vector Archive Mapping, Attentive Statistics Pooling, and residual connection, respectively.}
\resizebox{\columnwidth}{!}{
\begin{tabular}{@{}lc@{}}
\toprule
\textbf{Module Configuration} & \textbf{EER (\%)} \\ 
\midrule
w/o VAM (Transformer with Res. + ASP) & 8.856 \\ 
w/ VAM, but Transformer w/o Res. & 8.529 \\ 
w/o Transformer (VAM + ASP) & 8.352 \\
\textbf{Full TVAMSP (Transformer with Res. + VAM + ASP)} & \textbf{8.334} \\ 
\bottomrule
\end{tabular}
}
\label{tab:ablation}
\end{table}

\subsection{Ablation Study of the TVAMSP Module}
The results show that all components contribute to performance. Removing VAM causes the largest degradation among the ablations, increasing the EER from 8.334\% to 8.856\%. This indicates that the Vector Archive is the main contributor within TVAMSP. Removing the Transformer residual connection also degrades performance, yielding an EER of 8.529\%. Interestingly, VAM without the Transformer still achieves 8.352\%, suggesting that archive-based mapping accounts for most of the gain, while the Transformer mainly refines the mapping process. The full TVAMSP achieves the best result.

\section{Conclusion}
In this paper, we addressed the challenge of speaker verification from short utterances by proposing VAM-ECAPA, a system featuring the core TVAMSP module. Its key innovation, the Vector Archive, compensates for information scarcity by enriching frame-level features, achieving a 54.8\% relative error reduction on 1-second test segments compared to a strong baseline. While our system shows strong robustness on short utterances, we identified a clear trade-off. When the input is already long enough, the archive-based compensation may interfere with naturally rich features, and performance on utterances of 3 seconds or longer did not surpass the conventionally-trained baseline. Future work will aim to resolve this by designing an adaptive architecture that adjusts the contribution of the TVAMSP module based on input length. We also plan to introduce explicit supervision over the Vector Archive so that each archive captures a distinct aspect of speaker traits, making the library more expressive. Testing our system under noisy conditions and across different languages remains an important next step \cite{exp3}.


\section{Acknowledgement}
This work was supported in part by: the Institute of Information \& Communications Technology Planning \& Evaluation (IITP) grant funded by the Korean government (MSIT) under Grant No. RS-2019-II190079 for the Artificial Intelligence Graduate School Program at Korea University; the Institute of Information \& Communications Technology Planning \& Evaluation (IITP) grant funded by the Korean government (MSIT) under Grant No. RS-2025-02304828 for the Artificial Intelligence Star Fellowship Support Program to Nurture the Best Talents; the Institute of Information \& Communications Technology Planning \& Evaluation (IITP) grant funded by the Korean government (MSIT) under Grant No. RS-2025-25442867; the National Research Foundation of Korea (NRF) grant funded by the Korean government (MSIT) under Grant No. RS-2025-24535409; the Technology Development Program funded by the Ministry of SMEs and Startups (MSS, Korea) under Grant No. RS-2026-25534256; and the Supreme Prosecutor's Office Research Grant in 2026 (research title: Development of fake voice detection technology robust in new voice generation technology and speaker recognition).

\section{Generative AI Use Disclosure}
Generative AI tools were used only for language editing and grammar polishing. All scientific content, including the research idea, methodology, experiments, analysis, and conclusions, was produced and verified by the authors.
\bibliographystyle{IEEEtran}
\bibliography{mybib}

\end{document}